# Quasi-Two-Dimensional Drops

**Authors:** Tytti Kärki[1], Into Pääkkönen[1], Nikos Kyriakopoulos[1] and Jaakko V. I. Timonen[1]*

**Affiliations:**

[1]Department of Applied Physics, Aalto University School of Science, Puumiehenkuja 2, 02150 Espoo, Finland

*Corresponding author. Email: jaakko.timonen@aalto.fi

**Abstract:** Liquid drops are everywhere around us and important in numerous technological applications. Here, we demonstrate a quasi-two-dimensional (Q2D) analogy to the regular, often close to axisymmetric, three-dimensional (3D) drops. The Q2D drops are created by confining liquids between vertical walls, leading to formation of low aspect ratio capillary bridges that are deformed by gravity. When stationary, the Q2D drops adopt projected shapes that are analogous to 3D sessile drops, ranging from circular drops to puddles. When moving, the Q2D drops exhibit capillary and fluid mechanical behaviours analogous to 3D drops, including impacts and sliding on pseudo-surfaces. The Q2D drops also exhibit considerably more complex phenomena such as levitation, instabilities and pattern formation when subjected to external electric, magnetic and flow fields – all seen also in regular 3D drops. The presented 3D-Q2D analogy suggests that the diverse and often complicated phenomena observed in 3D drops can be studied in the Q2D geometry, allowing also new physics arising from the reduced dimensionality and the new boundary conditions.

**Main Text:** Sessile liquid drops are ubiquitous and diverse around us. Depending on the drop volume, properties of the liquid, and the nature of interaction with the supporting surface, the drops can adopt shapes ranging from ideal spheres and puddles to highly irregular geometries. It is well established how surface tension, contact angle (CA) and external forces such as gravity determine the overall equilibrium drop shape(*1*). The situation is especially interesting when the CA is large, leading to drops beading up and often moving nearly effortlessly on the surface(*2*, *3*). Such non-wetting is useful in numerous applications where dryness and self-cleaning are important(*4–6*). In practice, large CAs and non-wetting states can be achieved by taking advantage of the Cassie-Baxter state on superhydrophobic surfaces(*5*, *7*, *8*), the Leidenfrost effect(*9–11*), or utilizing the so called fizzy drops that were discovered recently(*12*, *13*).

      In this article, we show that there exists a quasi-two-dimensional (Q2D) analogy for the above-mentioned and widely studied three-dimensional (3D) sessile drops on non-wetting surfaces. The analogy is based on confining liquid between two flat and wetting parallel vertical walls with a narrow gap between, leading to a thin capillary bridge. We show that these flat vertical capillary bridges behave surprisingly similarly to 3D drops, showing analogous static and dynamic behaviors under a variety of external driving fields including gravity, shear, and electric and magnetic fields. (*14*).

**Formation and equilibrium shape of stationary Q2D liquid drops**

Formation of an ideal Q2D pseudo-sessile drops requires a sample cell (Hele-Shaw cell(*15*)) consisting of two rigid transparent substrates with an uniform gap between them. To achieve this, we used 4 mm thick optical glass windows with $\lambda/4$ flatness. These were thoroughly cleaned to suppress contact angle hysteresis and pinning and assembled together using plastic spacers and metal clips under HEPA filtered air to achieve an essentially dust free cell with a gap $b \sim 100$ μm between the confining surfaces (see Methods for details). While Q2D drops can be, in principle, prepared in a cell with any gap smaller than the capillary length, we chose to work in the range of $b \sim 100$ μm as therein the gravitational force ($F_g \sim Ab$, where $A$ is the projected area of the Q2D drop) is large enough to overcome the remaining (minor) contact line pinning ($F_{CL} \sim \sqrt{A}$), enabling close to ideal equilibrium drop shapes. The cell thickness was quantified from multiple points using white light interferometry for each constructed cell (**fig. S1**, **table S1**) with typical non-uniformity of $\Delta b \sim 1$ μm within the multicentimetre-sized cell.

The resemblance between the traditional 3D sessile drops encountered on non-wetting surfaces and the Q2D pseudo-sessile drops is obvious (**Fig. 1A,B**). While the 3D drops appear mostly dark (**Fig. 1A**), the Q2D drops are uniformly transparent with only a narrow dark band near the edge of the drop (**Fig. 1B**). In the 3D drops, the nearly entirely dark appearance is caused by refraction of the background light from the curved liquid-air interface (**Fig. 1C**). In contrast, refraction in the Q2D drops is present only near the drop edges where the concave meniscus is present (**Fig. 1D**), leading to overall even passage of light through the drop. With approximately collimated illumination and low numerical aperture imaging, the expected width of the dark boundary is ca. $b/2 \sim 50$ μm, as observed experimentally. The settled down Q2D drops appear to be sitting on a dark substrate (**Fig. 1B**) that does not reflect light as in the case of 3D drops to create a mirror image of the drop (**Fig. 1A**). This is because the apparent surface is not a real surface but rather a pseudo-surface that arises from the refraction of light from the edge of the glass window (**Fig. 1D**). In reality, the Q2D drops are sitting entirely on air. As for the 3D drops on non-wetting surfaces (**Fig. 1C**), the Q2D drops also exhibit an apparent contact angle ($\theta_E^*$) of ca. 180° (**Fig. 1D**).

A theoretical model describing the shape of the Q2D pseudo-sessile drops can be derived by modifying the widely used model for 3D drops(*1, 16*). The model describes the local drop half-width $r$ as a function of distance $z$ starting from the drop apex (**Fig. 1E**). Starting from the hydrostatic pressure $p(z) = \rho g z$ (where $\rho$ is liquid density and $g$ is gravitational acceleration) and the Laplace pressure $\Delta p = \gamma \left(\frac{1}{R_1} + \frac{1}{R_2}\right)$ (where $\gamma$ is surface tension and $R_1$ and $R_2$ are the two principal radii of curvature), the governing equation for the Q2D drop shape can be derived to be (see **Supplementary Information**)

$$p(z) - p_0 = \gamma \left(\frac{1}{R_{xz(z)}} + \frac{1}{R_{ny}}\right), \qquad \text{(Eq. 1)}$$

where $\frac{1}{R_{xz(z)}} = \nabla \hat{n} = \frac{r_{zz}}{(1+r_z^2)^{\frac{3}{2}}}$, where $r_z = \frac{dr}{dz}$ and $r_{zz} = \frac{d^2r}{dz^2}$. The main difference compared to the analogous governing equation for the 3D sessile drops is the slightly different curvature term (see **Supplementary Information**). However, both the 3D and Q2D governing equations lead to the same capillary length scale, $\lambda_c = \sqrt{\frac{\gamma}{\rho g}}$. Eq. (1) can be numerically solved to obtain $r(z)$, from which the drop width $w$, height $h$, and projected area $A$ can be extracted and compared to the experiments (see **Supplementary Information**). Perfect wetting with 0° local CA and hemispherical meniscus is assumed everywhere along the drop boundary (**Fig. 1F**). In the case of Q2D water drops, the true contact angle was confirmed to be < 3° using confocal reflection interference contract microscopy (RICM), as expected for water on clean glass (**Fig. 1F**, **fig. S2**).

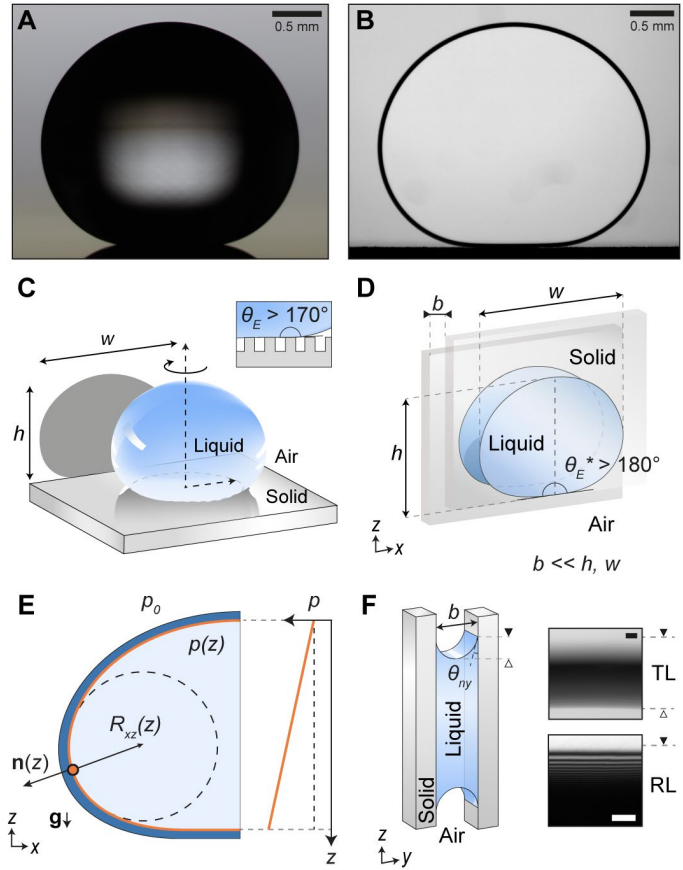

**Figure 1. Quasi-two-dimensional (Q2D) pseudo-sessile drops resemble three-dimensional (3D) sessile drops on non-wetting surfaces.** (**A**) An image of a 3D sessile water drop on a superhydrophobic surface based on micropillars and low surface energy coating. Image kindly provided by Sakari Lepikko. (**B**) An image of a Q2D pseudo-sessile water drop on a non-wetting pseudo-surface formed between two wetting glass windows ($b$ = 120 µm). (**C**) A scheme corresponding to the axisymmetric 3D drop in (A) with a large apparent contact angle ($\theta_E$). The inset shows the micropillar structure of the substrate. (**D**) A scheme corresponding to the Q2D drop in (B) viewed from the front. The Q2D drop has an apparent contact angle ($\theta_E^*$) of perfect 180°. (**E**) The model geometry for the equilibrium shape of the Q2D pseudo-sessile drops. Hydrostatic pressure $p(z)$ is balanced by Laplace pressure $p(z) - p_0$, where $p_0$ is pressure of the gas/vapor phase. (**F**) A scheme corresponding to the Q2D drop in (B) viewed from the side, where $\theta_{ny}$ is close to zero. The insets show transmitted and reflected light (TL, RL, respectively) microscopy images of the Q2D drop meniscus where interference fringes arise from the thin liquid wedge. Scale bars are 10 µm.

**Falling down, equilibrium shape and evaporation of Q2D drops**

The Q2D drops can be pipetted in from the bottom or top of the vertical sample cell, and in the latter case, Q2D drops fall within the confinement (**movie S1, Fig. 2A,B**) analogously to 3D drops falling in air(*17*). In contrast to the falling 3D drops, the terminal velocities of typical falling Q2D water drops with volume $V_i \sim 1$ µl are orders of magnitude slower ($\sim 10^{-3}$ m s$^{-1}$) and the Q2D drops leave behind a thin liquid film of thickness $e$ (**Fig. 2A,B**) similarly to liquid slugs moving in capillaries(*18, 19*). The slower falling velocity is caused by the significantly larger viscous dissipation due to the confinement and the no-slip boundary conditions. Because the kinetic energy is very modest in Q2D drops, the impact is also much gentler than in 3D(*20, 21*). This is apparent when comparing Weber numbers We $= \rho v^2 L/\gamma$ (where $L$ is the characteristic length scale) of millimetric 3D and Q2D water drops released at a height of 3 cm: Just before drop impacting the surface the Weber number is reduced from ca. We $\sim 9$ (3D) to ca. We $\sim 5 \times 10^{-5}$ (Q2D) due to the confinement-increased viscous dissipation that slows down the drops ($v_{3D} = 0.8$ m s$^{-1}$, $v_{Q2D} = 0.002$ m s$^{-1}$, $L_{3D} = L_{Q2D} = R$). After the impact, the Q2D drops adopt the equilibrium shape described earlier within few seconds and remain stationary thereafter.

We investigated experimentally the equilibrium shapes of the Q2D pseudo-sessile drops as a function of drop volume for three liquids with different capillary lengths: water ($\lambda_c = 2.7$ mm), silicone oil ($\lambda_c = 1.5$ mm) and perfluoropolyether (PFPE, $\lambda_c = 1.0$ mm) (**table S2**). For all studied liquids, both the drop height and width increased monotonically with the increasing drop volume (**Fig. 2C**). As with 3D drops, the Q2D drops became increasingly deformed by gravity with increasing drop volume and decreasing capillary length (**Fig. 2D**). The Q2D drop heights saturated for all liquids at high volume limit, analogous to the puddle limit known for 3D drops. At the Q2D puddle limit, the puddle heights were 4.70 mm, 2.55 mm and 1.75 mm for water, silicone oil and PFPE, respectively.

Comparison between the theoretically predicted drop shapes and the experimental ones was done by overlaying the predicted shape from the model on the experimental images and adjusting the apex curvature $k_0$ to match the experimental drop height. We noticed that the experimentally observed drop shapes were systematically slightly wider than those predicted theoretically (**Fig. 2E**). Interestingly, if the capillary length was relaxed to a free parameter, in addition to the apex curvature, essentially perfect fits between the experimental and the model drops were obtained for all liquids and drop volumes (**Fig. 2C,E**). This is especially notable at the Q2D puddle limit, at which the theory predicts the puddle height to be $2\lambda_c$, but experimental values were only 85–88% of this. This is very close to the effective capillary length obtained using Park-Homsy correction(*22*), i.e. $\lambda_{PH} = \sqrt{\pi/4}\,\lambda_c \approx 0.89\lambda_c$, which is associated to the boundary conditions in the Hele-Shaw geometry when the liquid wets the solid surface and the aspect ratio ($b/R$) is small, as in our experiments.

In contrast to 3D drops, the Q2D geometry allows a convenient way to adjust the strength of the gravitational force acting on the Q2D drops. This is done by adjusting the component of the gravitational force along the Q2D drop plane by tilting the cell along the x-axis (**Fig. 2F,G**). Tilting enables the tangential component of the gravitational acceleration to be varied from $g \sim 0$ m s$^{-2}$ to $g \sim 9.81$ m s$^{-2}$, hence allowing one to effectively control the gravity and observe a transition from gravity-deformed drops with aspect ratio $h/w <1$ to perfectly circular drops with $h/w = 1$ (**Fig. 2F,G**). We note that analogous experiments with 3D drops would be considerably more challenging and require utilization of additional forces such as centrifugal(*23*) or magnetic forces(*24*) or controlled micro-gravity environment available in drop tower facilities(*25, 26*), space(*27, 28*) and zero-g flights(*29*).

The evaporation of Q2D drops is expected to differ from the 3D counterparts because of the liquid shape and the boundary conditions. Using density-matched tracer particles (**Fig. 2I, movie S2**), we observed dipolar flow field inside an evaporating water Q2D drop with $u \sim 10^{-4}$ m s$^{-1}$ peripheral flow along the drop surface from the apex of the drop to the bottom. Countering flow at the middle of the drop upwards was observed. The interfacial flow is driven by surface tension gradient $\gamma(\varphi)$ arising from a temperature gradient (Marangoni flow) caused by faster evaporation of water near the bottom of the drop exposed to ambient humidity compared to the drop apex being in equilibrium with saturated water vapor. As Marangoni flow velocity scales as $u \sim \Delta\gamma/\eta$(REF), the magnitude of the surface tension difference between the bottom and the apex is $\sim 10^{-7}$ N m$^{-1}$.

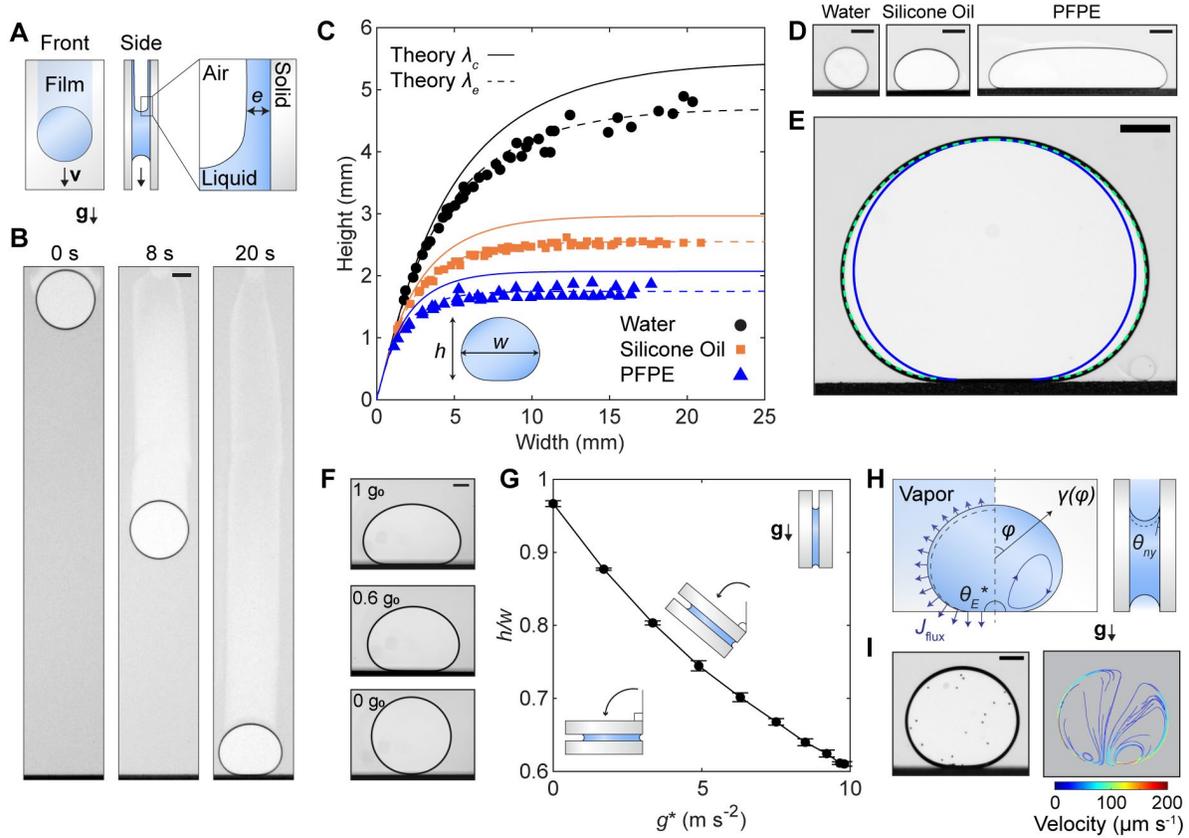

**Figure 2. Equilibrium shapes of the Q2D pseudo-sessile drops.** (**A**) A scheme showing front and side profiles of a falling Q2D drop that leaves a residual film on the glass surfaces. Close-up of the receding part of a falling Q2D drop, where a liquid film of thickness *e* is indicated. (**B**) Time series of a falling Q2D water drop ($b$ = 120 µm). Scale bar is 1 mm. (**C**) Height *h* of the Q2D pseudo-sessile drops as a function drop width *w*. Black circles, red squares, and blue triangles represent experimental data for water, silicone oil and PFPE, respectively. Solid and dashed lines are theoretical fits using capillary length ($\lambda_c$) and effective capillary length ($\lambda_e$), respectively. (**D**) Images of water, silicone oil and PFPE Q2D pseudo-sessile drops of same height (1.7 mm). All scale bars are 1 mm. (**E**) An image of a Q2D pseudo-sessile water drop overlayed with theoretical fits using $\lambda_c$ = 2.73 mm (solid orange line) and $\lambda_e$ = 2.26 mm (dashed orange line). Scale bar is 0.5 mm. (**F**) Images of a Q2D pseudo-sessile silicone oil drop under effective gravity of 9.81 m s$^{-2}$, 6 and 0 m s$^{-2}$. (**G**) Q2D silicone oil drop aspect ratio *h/w* as a function of effective gravity. The error bars are std for three experiments. (**H**) Schemes for an evaporating Q2D drop viewed from the front and the side. Higher evaporative flux at the open sample cell bottom induces surface tension gradient $\gamma(\varphi)$ that drive Marangoni flows. (**I**) An image of Q2D water drop with PE particles and corresponding time-lapse particle trajectories and their velocities measured during 2.5 min time span. Scale bar is 0.5 mm.

**Dynamics of the Q2D liquid drops: Sliding on tilted pseudo-surfaces, flow fields and contact line effects**

In addition to the falling dynamics and impacts, the Q2D drops can move laterally on the pseudo-surfaces when the cell is tilted. (**Fig. 3A**). This is analogous to 3D drops moving on tilted surfaces, which is often used to probe the droplet interaction with the underlying substrate and to assess quality of various surface coatings(*30–33*). The laterally moving Q2D drops leave a thin liquid film behind similarly as in the drop falling experiment (**Fig. 2A,B, Fig. 3A**). We observed that the Q2D pseudo-sessile drops began to move at very small tilting angles $\beta$ (**Fig. 3B**), indicating that lateral adhesion forces are small(*34–37*). The critical tilting angle ($\beta_c$) required to make a Q2D PFPE drop move was observed to be only 0.2–0.5°. Similarly, the typical terminal velocities were very small too, approximately 0.01 mm s$^{-1}$. This is in agreement with viscosity-limited models developed for capillary bridges(*38–41*). Analogous to 3D drops where the terminal velocity scales as $\gamma \sin\beta/\eta$(*3, 42*), the velocity of the PFPE Q2D drops increased linearly with the tilt angle (**Fig. 3B**). In addition, a decrease of the velocity with increasing viscosity was observed by using water-glycerol mixtures (**Fig. 3C**). These observations suggest that the viscous dissipation can still dominate over other frictional forces such as contact line pinning, despite that the length of the three-phase contact line in Q2D drops is considerably larger than for the 3D counterparts.

Finally, we analysed the flow fields inside moving Q2D pseudo-sessile drops using tracer particles. In contrast to large CA 3D drops that can move by sliding or rolling(*32, 43, 44*), the Q2D drops appear to move solely by sliding. This is caused by the vertical no-slip boundaries at the two sides of the drop. As Reynolds number $Re = \rho v L/\eta$ in moving Q2D drops is small (e.g. Re = 0.2 for a Q2D water drop sliding with 1 mm s$^{-1}$ velocity when $L$ = 220 µm), the flow is laminar and parabolic between these two confining surfaces(*40*) (**Fig. 3D**). This was investigated using silicone oil Q2D drops containing PE particles. The drop ($V \sim 2$ µl) moving with a velocity of $\sim 0.1$ mm s$^{-1}$ (Re <10$^{-3}$) was observed to contain particles moving at velocities ranging from 0 mm s$^{-1}$ to 0.23 mm s$^{-1}$ (**Fig. 3E**) with highest concentration of particles moving at 0.15 mm s$^{-1}$. This peak could be explained by parabolic profile and the particles preferring to localize at the distance of 0.6$b$/2 away from the midplane, known as the Segre-Silberberg effect(*45, 46*).

The experiments with tracer particles also revealed, unexpectedly, more complex behaviors for some liquids like water, where backwards directed flows near the perimeter of the drop were observed (**movie S3**, **Fig. 3F**). This contrasts with the silicone oil experiments where such backwards flows were absent. The mean velocities of Q2D water drops were also observed to be notably smaller (down to 80% smaller) than expected based on viscous dissipation only (**Fig. 3G**). This suggests a presence of additional friction forces, likely acting on the contact line or in the precursor film where dynamics can be very different for non-volatile viscous oils (pseudo-partial wetting(*47*)) compared to water where evaporation occurs continuously at the drop interfaces and from the film.

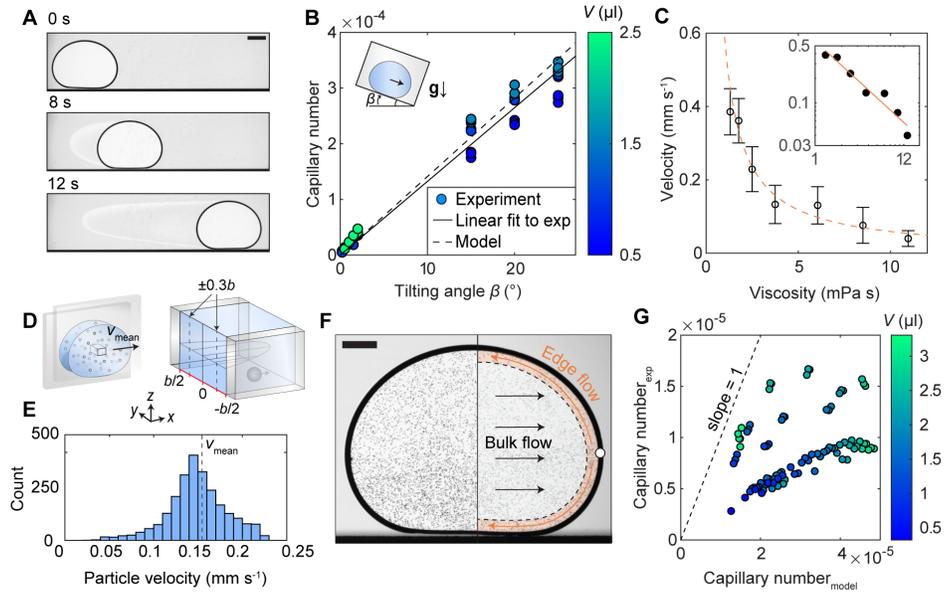

**Figure 3. Q2D drops moving on tilted pseudo-surfaces.** (**A**) Time series of a Q2D water drop moving on a pseudo-surface ($\beta = 15°$, scale bars are 1 mm). (**B**) Capillary number ($Ca$) as a function of $\beta$ for Q2D PFPE drops. Filled circles correspond to experimental data, and solid and dashed lines are linear fits to the experimental data and the theoretical model, respectively. Drop volume is indicated by the colormap. (**C**) Q2D drop velocity as a function of drop viscosity for water-glycerol mixtures ($V \sim 1.5$ µl, $\beta = 10°$). Circles correspond to experimental mean velocities from 5 to 23 drops each. Error bars indicate the standard deviation. Dashed orange line corresponds to best fit of $v \propto 1/\eta$. Inset shows the same data in a log-log plot. (**D**) Scheme of a laterally moving Q2D drop containing tracer particles and magnified view of the expected flow field with the Segre-Silberberg distance/location indicated. (**E**) Velocity histogram of measured particle velocities in laterally moving Q2D silicone oil drop. (**F**) Snapshot of a moving Q2D pseudo-sessile water drop with polystyrene particles (**movie S3**) and a simplified overlayed scheme of the observed flow patterns: Bulk flow towards the direction of motion (black arrows) and countering edge flows (orange arrows). The white circle indicated the stagnation point. Scale bar is 0.5 mm. (**G**) Experimental Ca as a function of model Ca for water Q2D drops with $\beta$ varying between 10 and 25°. $b = 220$ µm for (a–g).

**More complex phenomena: Levitation, magnetic instabilities and electrohydrodynamic structuring**

Finally, in addition to the falling and sliding dynamics driven by gravity, we tested further whether other more complex phenomena seen in 3D sessile drops driven by shear(*9, 10, 48*), magnetic(*49–51*) and electric fields(*52*) can be seen in Q2D drops. Firstly, 3D drops moving relative to the surrounding gas phase are known to experience shear force that can lead to complex drop shapes, fragmentation and even levitation(*17, 53–57*). Analogously in Q2D drops, aerodynamic lift and levitation can be achieved by overcoming the gravitational force $F_G$ with aerodynamic drag force $F_D$ created using upwards gas flow in the Hele-Shaw cell (**Fig. 4A,B, movie S4**). The levitating Q2D drops can also undergo fragmentation, especially when the surface tension is small (**movie S5, Fig. 4C**). This is analogous to the falling large 3D drops that disintegrate during the fall, limiting for example the size of the raindrops falling from the sky(*17, 53, 54*).

Secondly, 3D drops of magnetic liquids have been known since 1960s to exhibit a variety of deformations and pattern formations in magnetic fields driven by reduction in magnetostatic energy(*50, 51*). Similarly, Q2D droplets made of a strongly magnetic colloidal dispersion, a ferrofluid, are expected to deform in external magnetic field (**Fig. 4D**). Indeed, when a magnetizable Q2D drop was subjected to a non-uniform magnetic field from a permanent magnet below the Q2D drop, we observed deformation of the drop followed by instability on the drop surface reminiscent of the well-known Rosensweig instability(*49–51*) (**Fig. 4E,F, movie S6**).

Thirdly, 3D drops are known to undergo a plethora of non-equilibrium structuring events when subjected to electric fields that drive electrohydrodynamic forces(*58–61*). Similarly, we observed that a Q2D drop of insulating perfluoropolyether (PFPE, $\sigma_{PFPE} \sim 10^{-17}$ S m$^{-1}$) surrounded by a slightly conductive mixture of Aerosol OT and dodecane ($\sigma_{AOT-DD} \sim 10^{-8}$ S m$^{-1}$)(*52*) confined between transparent indium tin oxide (ITO) coated glass slides providing perpendicular electric field(*52, 62*) can undergo various complicated electrohydrodynamically driven structuring events (**Fig. 4G,H, movie S7**).

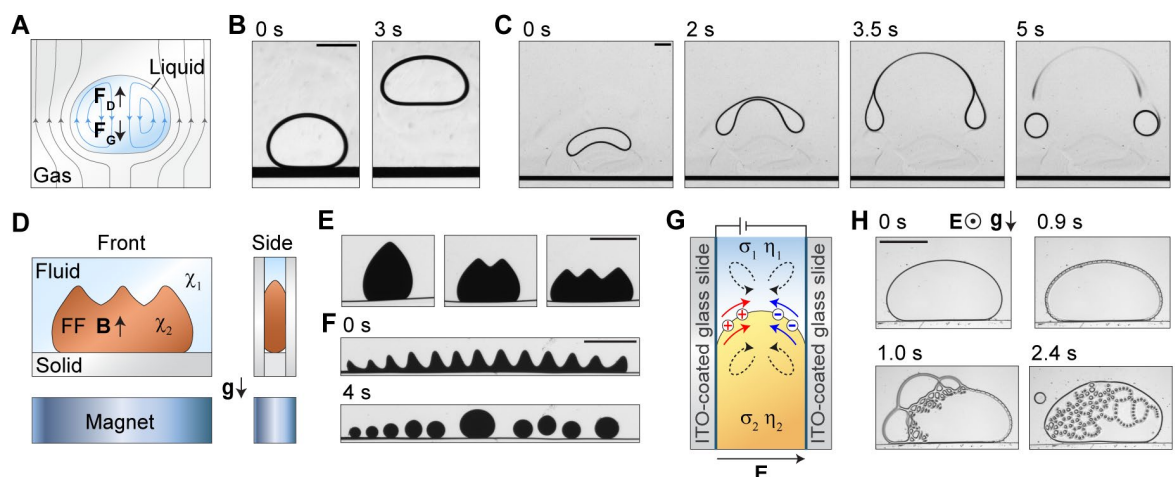

**Figure 4. Levitation, magnetic instabilities and electrohydrodynamic structuring in Q2D pseudo-sessile drops.** (**A**) Schematic front view of a Q2D drop levitating under in-plane gas flow. (**B**) Time series of a small Q2D PFPE drop levitating under planar N$_2$ flow ($V \sim 0.5$ μl; $b = 220$ μm). (**C**) Time series of a larger Q2D PFPE drop levitating and splitting under planar N$_2$ flow (initial drop $V \sim 1$ μl; $b = 220$ μm). (**D**) Schematic front and side views of a magnetizable Q2D drop (ferrofluid, FF) undergoing instabilities in applied nonhomogeneous magnetic field created by the permanent magnet under the Hele-Shaw cell. (**E**) Three microscopy images showing a magnetizable fluorocarbon based Q2D drop surrounded by solution of dodecane containing Aerosol OT surfactant in three states of increasing magnetic field strength ($V \sim 0.03$ μl; $b \sim 60$ μm). (**F**) Time series of the magnetizable Q2D drop undergoing instabilities that result in splitting of the drop ($V_i \sim 0.05$ μl; $b \sim 60$ μm). (**G**) Schematic side view of the interface between a Q2D drop and a surrounding liquid in an electric field. (**H**) Time series of microscopy images of a PFPE Q2D drop undergoing electrohydrodynamically driven structuring ($V \sim 0.05$ μl; $b \sim 60$ μm). All scale bars are 1 mm.

**Outlook**

We have realized quasi-two-dimensional (Q2D) liquid drops with large apparent contact angles as an analogy for the widely studied three-dimensional (3D) drops on non-wetting surfaces. Many of the static and dynamic behaviours of the Q2D drops in gravitational, shear, electric and magnetic field are conceptually analogous to those of the conventional 3D drops studied on superhydrophobic surfaces(*2*, *4–6*, *8*, *32*, *62*), superoleophobic surfaces(*63*, *64*), Leidenfrost states(*9*, *10*, *65*) and as fizzy drops(*12*, *13*). We foresee that this analogy, combined to the excellent optical access to the droplet contents and the possibility of observing new physics arising from the reduced dimensionality and boundary conditions, will lead to discovery of various novel capillary phenomena not possible in conventional 3D drops and also to new insight to open fundamental questions related to e.g. droplet friction(*66*).

**Materials and Methods**

Materials

*Sample cells.* Optical windows (Edmund Optics, uncoated square borosilicate-crown glass, 50 x 50 mm, $\lambda/4$ surface flatness 47-944), glass objective slides (VWR 631-1552), indium tin oxide (ITO) coated glass slides (nominal thickness of coating ~350 nm; Diamonds Coatings Ltd.), lens cleaning tissues (Thorlabs MC-50E), plastic spacers (PrecisionBrand, plastic shim set 77-644-905), thermoplastic Surlyn ionomer film (DuPont 1702) and office clippers.

*Liquids.* Ethanol (Altia 1025874)**,** silicone oil (Sigma-Aldrich 317667), perfluoropolyether (Chemours Performance Lubricants GPL 102-500), glycerol (Fisher BioReagents BP2291) and ferrofluid (Ferrotec DFF2, lot. T041019A) were used as received. Distilled (DI) water (MerckMillipore Direct-Q® 3 UV Water Purification System ZRQSVP3WW) was filtered (Fisher, 0.2-0.45 µm pore size 15206869, 17144381) before experiments. See **Table S2** for details. 150 mM docusate sodium salt (AOT) solution in *n*-dodecane was prepared by dissolving AOT (Sigma-Aldrich, 99%) in *n*-dodecane (Acros Organics, 99%, anhydrous).

*Particles.* Grey Polyethylene Microspheres (PE, Cospheric, density 1.00g/cc, diameter distribution 10–45 µm GRYPMS-1.00) and polystyrene microparticles (MicroParticles GmbH, 10 µm, PS/Q-R-KM544) were either used as received or washed by centrifugation, respectively. The density-matching of the PE particles was done by centrifugation, followed by discarding the sedimented and creamed particles as done before(*67*). Further, the particle dispersion was left to sediment/cream overnight, after which the particle dispersion from the middle part of the container was used for the experiments.

Methods

All experiments were performed under ambient laboratory conditions.

*Preparation of Q2D sample cells*. Optical glass windows were cleaned by pipetting a ca. 50 µl solvent drop on the glass surface, placing a lens cleaning tissue on the drop and dragging the wet tissue slowly on the surface, leaving behind a dry surface. This was done first with DI water and then with ethanol, ca. three times in different directions with both liquids. Additionally, optical window edges were cleaned with ethanol-wetted cotton swabs. Lastly the windows were rinsed with DI water and dried with a nitrogen gun. Subsequently, the optical windows were plasma activated (Henniker Scientific Ltd. HPT-100, air plasma, 80% power, 3 minutes), and the cell was assembled in a HEPA-filtered biological safety cabinet (Kojair Biowizard Golden Line) using 102 or 191 µm thick and 1–2 cm wide plastic spacers that were wiped with ethanol and dried before sandwiching between the optical windows. Finally, the optical windows and spacers were clamped together with office clippers. The sample cell was protected from airborne dust at all times when possible. Just prior to the experiments, a nitrogen gun was used to remove any accumulated dust from the cell.

***Optical imaging.*** Custom imaging setups consisted of a 0.16x or 1x objective (EdmundOptics 56-675, Mitutoyo 378-800-3) coupled either directly, or via tube lens (Thorlabs), to a USB-camera (Ximea MC050CG-SY, Basler 35-927, FLIR GS3-U3-51S5M-C). The light source in the custom set ups was either a LED panel (Godox LEDP120C, 45939) or collimated LED (Thorlabs MWWHLP1, SM2F32-A). Additional imaging was done with a commercial goniometer (Biolin Scientific Attention Theta).

***Preparation and manipulation of Q2D drops.*** The Q2D sample cell was placed on a sample stage composed of a plate holder (Thorlabs FP02) mounted to a tilting stage (Thorlabs GNL10/M) and xyz-translation stage (Thorlabs PT3/M). The tilting stage was used to align the sample cell normal to gravity. Experiments started with a small liquid volume (0.5–1 µl) pipetted into the sample cell, followed by drop size being increased stepwise by adding ~0.5–1 µl of the liquid and acquiring a new image after each addition. The effect of the tangential component of the gravity on the equilibrium shape of the Q2D drop was studied by placing the Q2D sample cell in a cage-system (Thorlabs) mounted on a goniometer stage (Thorlabs NR360S/M) that allowed the tilting of the sample cell 0–90° with respect to the horizon. On the other hand, to observe falling Q2D drops, the liquid ($V$ ~1 µl) was pipetted in a single step from the top of the sample cell and the falling of the Q2D drop was recorded. Finally, drop sliding experiments were done with a custom setup consisting of a tilting stage (Thorlabs GNL10/M) for tilt angles $\beta \leq 10°$, and using a commercial goniometer (Biolin Scientific Attention Theta) for tilt angles $\beta > 10°$.

***Levitation, magnetic instabilities and electrohydrodynamic structuring in Q2D drops.*** For Q2D drop levitation experiments, the sample cell and Q2D drop was constructed from optical windows as described earlier in '*Preparation of Q2D sample cells*'. A Q2D PFPE drop in the bottom of a vertical sample cell was exposed to gas flow with a manually operated $N_2$ gun positioned below the sample cell. The ferrofluid and EHD experiments were done in a vertical sample cell constructed as described earlier(*52, 62*). The cell was first filled with AOT-DD (leaving some room for the Q2D drop), and then, either ferrofluid or PFPE was pipetted inside the sample cell creating a Q2D drop that was let to sediment on the bottom of the vertical sealed sample cell. In the ferrofluid Q2D drop experiments, a permanent cylindrical NdFeB magnet (1" diameter, 1" height, K&J Magnetics DX0X0-N52) was slowly moved beneath the sample cell. The distance between the bottom of the ferrofluid drop and magnet varied from 5 mm (B-field ~0.3 T) to 15 cm (0 T). In the EHD experiments, an electrometer (Keysight B2987A) was connected to the ITO-coated glass sample cell with wires (Alpha Wire 2936), and a DC electric field of magnitude of 4.14 V/µm was applied for 250 ms and then lowered to 3.5 V/µm until the electric field was switched off after ~10 s.

***Measurement of Q2D drop dimensions, area, and position.*** Drop height ($h$) and width ($w$) were measured from the inner edges of the Q2D drops using the FIJI/ImageJ(*68, 69*) profile plot tool. In addition, a custom MATLAB script was used to detect the boundaries of the Q2D drops and thus obtain the drop profile $r(z)$, area $A$ and volume $V = Ab$. The position of a moving Q2D drop was measured with another custom MATLAB script.

***Tracking and measurement of particle velocities in Q2D drops.*** The experimental videos were cropped with moving region of interest using a custom MATLAB code and processed (threshold, binarization) prior to analysis performed with FIJI/ImageJ(*68, 69*) plug-in TrackMate(*70, 71*).

# References

1. P. G. de Gennes, F. Brochard-Wyart, D. Quéré, *Capillarity and Wetting Phenomena: Drops, Bubbles, Pearls, Waves* (Springer, New York, NY, translation of the French ed., 2010).
2. A. Lafuma, D. Quéré, Superhydrophobic states. *Nat. Mater.* **2**, 457–460 (2003).
3. T. Mouterde, P. S. Raux, C. Clanet, D. Quéré, Superhydrophobic frictions. *Proc. Natl. Acad. Sci. U.S.A.* **116**, 8220–8223 (2019).
4. D. Wang, Q. Sun, M. J. Hokkanen, C. Zhang, F.-Y. Lin, Q. Liu, S.-P. Zhu, T. Zhou, Q. Chang, B. He, Q. Zhou, L. Chen, Z. Wang, R. H. A. Ras, X. Deng, Design of robust superhydrophobic surfaces. *Nature* **582**, 55–59 (2020).
5. D. Quéré, M. Reyssat, Non-adhesive lotus and other hydrophobic materials. *Phil. Trans. R. Soc. A.* **366**, 1539–1556 (2008).
6. X. Tian, T. Verho, R. H. A. Ras, Moving superhydrophobic surfaces toward real-world applications. *Science* **352**, 142–143 (2016).
7. A. B. D. Cassie, S. Baxter, Wettability of porous surfaces. *Trans. Faraday Soc.* **40**, 546 (1944).
8. D. Quéré, Wetting and roughness. *Annu. Rev. Mater. Res.* **38**, 71–99 (2008).
9. A.-L. Biance, C. Clanet, D. Quéré, Leidenfrost drops. *Phys. Fluids* **15**, 1632 (2003).
10. D. Quéré, Leidenfrost dynamics. *Annu. Rev. Fluid Mech.* **45**, 197–215 (2013).
11. G. P. Neitzel, P. Dell'Aversana, Noncolaescence and nonwetting behavior of liquids. *Annu. Rev. Fluid Mech.* **34**, 267–289 (2002).
12. D. Panchanathan, P. Bourrianne, P. Nicollier, A. Chottratanapituk, K. K. Varanasi, G. H. McKinley, Levitation of fizzy drops. *Sci. Adv.* **7**, eabf0888 (2021).
13. M. Backholm, T. Kärki, H. Nurmi, M. Vuckovac, V. Turkki, S. Lepikko, V. Jokinen, J. V. I. Timonen, R. H. A. Ras, Drop dissipation on superhydrophobic surfaces at the limit of vanishing and zero contact-line friction. Submitted (Sci. Adv.).
14. X. Li, F. Bodziony, M. Yin, H. Marschall, R. Berger, H.-J. Butt, Kinetic drop friction. *Nat. Commun.* **14**, 4571 (2023).
15. H. S. Hele-Shaw, The flow of water. *Nature* **58**, 34–36 (1898).
16. S. M. I. Saad, A. W. Neumann, Axisymmetric drop shape analysis (ADSA): An outline. *Adv. Colloid Interface Sci.* **238**, 62–87 (2016).
17. E. Villermaux, B. Bossa, Single-drop fragmentation determines size distribution of raindrops. *Nat. Phys.* **5**, 697–702 (2009).
18. J. Bico, D. Quéré, Falling slugs. *J. Colloid Interface Sci.* **243**, 262–264 (2001).
19. O. E. Jensen, Draining collars and lenses in liquid-lined vertical tubes. *J. Colloid Interface Sci.* **221**, 38–49 (2000).
20. D. Khojasteh, M. Kazerooni, S. Salarian, R. Kamali, Droplet impact on superhydrophobic surfaces: A review of recent developments. *J. Ind. Eng. Chem.* **42**, 1–14 (2016).
21. A. L. Yarin, Drop impact dynamics: Splashing, spreading, receding, bouncing…. *Annu. Rev. Fluid Mech.* **38**, 159–192 (2006).
22. C.-W. Park, G. M. Homsy, Two-phase displacement in Hele Shaw cells: Theory. *J. Fluid Mech.* **139**, 291–308 (1984).
23. R. Tadmor, P. Bahadur, A. Leh, H. E. N'guessan, R. Jaini, L. Dang, Measurement of Lateral Adhesion Forces at the Interface between a Liquid Drop and a Substrate. *Phys. Rev. Lett.* **103**, 266101 (2009).
24. J. V. I. Timonen, M. Latikka, O. Ikkala, R. H. A. Ras, Free-decay and resonant methods for investigating the fundamental limit of superhydrophobicity. *Nat. Commun.* **4**, 2398 (2013).
25. A. Diana, M. Castillo, D. Brutin, T. Steinberg, Sessile Drop Wettability in Normal and Reduced Gravity. *Microgravity Sci. Technol.* **24**, 195–202 (2012).
26. Z.-Q. Zhu, D. Brutin, Q.-S. Liu, Y. Wang, A. Mourembles, J.-C. Xie, L. Tadrist, Experimental Investigation of Pendant and Sessile Drops in Microgravity. *Microgravity Sci. Technol.* **22**, 339–345 (2010).
27. S. Kumar, M. Medale, P. D. Marco, D. Brutin, Sessile volatile drop evaporation under microgravity. *NPJ Microgravity* **6**, 37 (2020).
28. J. McCraney, V. Kern, J. B. Bostwick, S. Daniel, P. H. Steen, Oscillations of drops with mobile contact lines on the international space station: Elucidation of terrestrial inertial droplet spreading. *Phys. Rev. Lett.* **129**, 084501 (2022).
29. D. Brutin, Z. Zhu, O. Rahli, J. Xie, Q. Liu, L. Tadrist, Sessile drop in microgravity: Creation, contact angle and interface. *Microgravity Sci. Technol.* **21**, 67–76 (2009).

30. H.-J. Butt, J. Liu, K. Koynov, B. Straub, C. Hinduja, I. Roismann, R. Berger, X. Li, D. Vollmer, W. Steffen, M. Kappl, Contact angle hysteresis. *Curr. Opin. Colloid Interface Sci.* **59**, 101574 (2022).
31. P. Olin, S. B. Lindström, T. Pettersson, L. Wågberg, Water drop friction on superhydrophobic surfaces. *Langmuir* **29**, 9079–9089 (2013).
32. B. S. Yilbas, A. Al-Sharafi, H. Ali, N. Al-Aqeeli, Dynamics of a water droplet on a hydrophobic inclined surface: influence of droplet size and surface inclination angle on droplet rolling. *RSC Adv.* **7**, 48806–48818 (2017).
33. N. Gao, F. Geyer, D. W. Pilat, S. Wooh, D. Vollmer, H.-J. Butt, R. Berger, How drops start sliding over solid surfaces. *Nat. Phys.* **14**, 191–196 (2018).
34. C. W. Extrand, Y. Kumagai, Liquid drops on an inclined plane: The relation between contact angles, drop shape, and retentive force. *J. Colloid Interface Sci.* **170**, 515–521 (1995).
35. D. W. Pilat, P. Papadopoulos, D. Schäffel, D. Vollmer, R. Berger, H.-J. Butt, Dynamic measurement of the force required to move a liquid drop on a solid surface. *Langmuir* **28**, 16812–16820 (2012).
36. A. I. ElSherbini, A. M. Jacobi, Retention forces and contact angles for critical liquid drops on non-horizontal surfaces. *J. Colloid Interface Sci.* **299**, 841–849 (2006).
37. M. Backholm, D. Molpeceres, M. Vuckovac, H. Nurmi, M. J. Hokkanen, V. Jokinen, J. V. I. Timonen, R. H. A. Ras, Water droplet friction and rolling dynamics on superhydrophobic surfaces. *Commun. Mater.* **1**, 64 (2020).
38. P. G. de Gennes, Wetting: Statics and dynamics. *Rev. Mod. Phys.* **57**, 827–863 (1985).
39. J. Eggers, H. A. Stone, Characteristic lengths at moving contact lines for a perfectly wetting fluid: The influence of speed on the dynamic contact angle. *J. Fluid Mech.* **505**, 309–321 (2004).
40. T. D. Blake, Slip between a liquid and a solid: D.M. Tolstoi's (1952) theory reconsidered. *Colloids Surf.* **47**, 135–145 (1990).
41. E. Reyssat, Capillary bridges between a plane and a cylindrical wall. *J. Fluid Mech.* **773**, R1 (2015).
42. E. Reyssat, Drops and bubbles in wedges. *J. Fluid Mech.* **748**, 641–662 (2014).
43. L. Mahadevan, Y. Pomeau, Rolling droplets. *Phys. Fluids* **11**, 2449–2453 (1999).
44. D. Richard, D. Quéré, Viscous drops rolling on a tilted non-wettable solid. *Europhys. Lett.* **48**, 286–291 (1999).
45. G. Segré, A. Silberberg, Radial Particle Displacements in Poiseuille Flow of Suspensions. *Nature* **189**, 209–210 (1961).
46. D. Huang, J. Man, D. Jiang, J. Zhao, N. Xiang, Inertial microfluidics: Recent advances. *Electrophoresis* **41**, 2166–2187 (2020).
47. F. Brochard-Wyart, J. M. Di Meglio, D. Quere, P. G. de Gennes, Spreading of nonvolatile liquids in a continuum picture. *Langmuir* **7**, 335–338 (1991).
48. G. Graeber, K. Regulagadda, P. Hodel, C. Küttel, D. Landolf, T. M. Schutzius, D. Poulikakos, Leidenfrost droplet trampolining. *Nat. Commun.* **12**, 1727 (2021).
49. R. E. Rosensweig, *Ferrohydrodynamics* (Dover Publications, Inc, Mineola, New York, Dover edition., 2014).
50. J. V. I. Timonen, M. Latikka, L. Leibler, R. H. A. Ras, O. Ikkala, Switchable static and dynamic self-assembly of magnetic droplets on superhydrophobic surfaces. *Science* **341**, 253–257 (2013).
51. M. Latikka, M. Backholm, A. Baidya, A. Ballesio, A. Serve, G. Beaune, J. V. I. Timonen, T. Pradeep, R. H. A. Ras, Ferrofluid microdroplet splitting for population-based microfluidics and interfacial tensiometry. *Adv. Sci.* **7**, 2000359 (2020).
52. G. Raju, N. Kyriakopoulos, J. V. I. Timonen, Diversity of non-equilibrium patterns and emergence of activity in confined electrohydrodynamically driven liquids. *Sci. Adv.* **7**, eabh1642 (2021).
53. W. R. Lane, Shatter of drops in streams of air. *Ind. Eng. Chem.* **43**, 1312–1317 (1951).
54. É. Reyssat, F. Chevy, A.-L. Biance, L. Petitjean, D. Quéré, Shape and instability of free-falling liquid globules. *Europhys. Lett.* **80**, 34005 (2007).
55. M. Jain, R. S. Prakash, G. Tomar, R. V. Ravikrishna, Secondary breakup of a drop at moderate Weber numbers. *Proc. R. Soc. A.* **471**, 20140930 (2015).
56. H. Zhao, D. Nguyen, D. Edgington-Mitchell, J. Soria, H.-F. Liu, D. Honnery, The largest diameter of falling drop in the up-gas flow. *Int. J. Multiph. Flow* **159**, 104345 (2023).
57. E. Villermaux, Fragmentation. *Annu. Rev. Fluid Mech.* **39**, 419–446 (2007).
58. G. I. Taylor, Studies in electrohydrodynamics. I. The circulation produced in a drop by an electric field. *Proc. R. Soc. Lond. A* **291**, 159–166 (1966).
59. J. R. Melcher, G. I. Taylor, Electrohydrodynamics: A review of the role of interfacial shear stresses. *Annu. Rev. Fluid Mech.* **1**, 111–146 (1969).

60. P. F. Salipante, P. M. Vlahovska, Electrohydrodynamics of drops in strong uniform dc electric fields. *Physics of Fluids* **22**, 112110 (2010).
61. P. M. Vlahovska, Electrohydrodynamics of drops and vesicles. *Annu. Rev. Fluid Mech.* **51**, 305–330 (2019).
62. R. Reyes Garza, N. Kyriakopoulos, Z. M. Cenev, C. Rigoni, J. V. I. Timonen, Magnetic Quincke rollers with tunable single-particle dynamics and collective states. *Sci. Adv.* **9**, eadh2522 (2023).
63. T. Verho, C. Bower, P. Andrew, S. Franssila, O. Ikkala, R. H. A. Ras, Mechanically durable superhydrophobic surfaces. *Adv. Mater.* **23**, 673–678 (2011).
64. A. Tuteja, W. Choi, M. Ma, J. M. Mabry, S. A. Mazzella, G. C. Rutledge, G. H. McKinley, R. E. Cohen, Designing superoleophobic surfaces. *Science* **318**, 1618–1622 (2007).
65. H.-J. Butt, C. Semprebon, P. Papadopoulos, D. Vollmer, M. Brinkmann, M. Ciccotti, Design principles for superamphiphobic surfaces. *Soft Matter* **9**, 418–428 (2013).
66. A. Bouillant, T. Mouterde, P. Bourrianne, A. Lagarde, C. Clanet, D. Quéré, Leidenfrost wheels. *Nat. Phys.* **14**, 1188–1192 (2018).
67. M. Vuckovac, M. Backholm, J. V. I. Timonen, R. H. A. Ras, Viscosity-enhanced droplet motion in sealed superhydrophobic capillaries. *Sci. Adv.* **6**, eaba5197 (2020).
68. J. Schindelin, I. Arganda-Carreras, E. Frise, V. Kaynig, M. Longair, T. Pietzsch, S. Preibisch, C. Rueden, S. Saalfeld, B. Schmid, J.-Y. Tinevez, D. J. White, V. Hartenstein, K. Eliceiri, P. Tomancak, A. Cardona, Fiji: an open-source platform for biological-image analysis. *Nat. Methods* **9**, 676–682 (2012).
69. C. A. Schneider, W. S. Rasband, K. W. Eliceiri, NIH Image to ImageJ: 25 years of image analysis. *Nat. Methods* **9**, 671–675 (2012).
70. J.-Y. Tinevez, N. Perry, J. Schindelin, G. M. Hoopes, G. D. Reynolds, E. Laplantine, S. Y. Bednarek, S. L. Shorte, K. W. Eliceiri, TrackMate: An open and extensible platform for single-particle tracking. *Methods* **115**, 80–90 (2017).
71. D. Ershov, M.-S. Phan, J. W. Pylvänäinen, S. U. Rigaud, L. Le Blanc, A. Charles-Orszag, J. R. W. Conway, R. F. Laine, N. H. Roy, D. Bonazzi, G. Duménil, G. Jacquemet, J.-Y. Tinevez, TrackMate 7: integrating state-of-the-art segmentation algorithms into tracking pipelines. *Nat. Methods* **19**, 829–832 (2022).
72. T. H. Besseling, J. Jose, A. V. Blaaderen, Methods to calibrate and scale axial distances in confocal microscopy as a function of refractive index. *J. Microsc.* **257**, 142–150 (2015).
73. A. Perot, C. Fabry, On the application of interference phenomena to the solution of various problems of spectroscopy and metrology. *ApJ.* **9**, 87 (1899).
74. D. Daniel, J. V. I. Timonen, R. Li, S. J. Velling, J. Aizenberg, Oleoplaning droplets on lubricated surfaces. *Nat. Phys. 13*, 1020–1025 (2017).

**Acknowledgments:** We thank Prof. Matilda Backholm and Ricardo Reyes Garza for valuable feedback and fruitful discussions.

**Funding:** Academy of Finland Center of Excellence Program (2022-2029) in Life-Inspired Hybrid Materials LIBER (346112) for J.V.I.T. Personal grant from Foundation for Aalto University Science and Technology for T.K.

**Competing interests:** Authors declare that they have no competing interests.